\documentclass[aps,pra,twocolumn,a4paper,10pt]{revtex4-2}
\usepackage{graphicx}
\usepackage{amsmath}
\usepackage{amssymb}
\usepackage{mathrsfs}
\usepackage{amsfonts}
\usepackage{dsfont}
\usepackage{dcolumn}
\usepackage{bm}
\usepackage{booktabs}
\usepackage{color}
\usepackage{xcolor}
\usepackage{colortbl}
\usepackage{mathtools}
\usepackage{upgreek}
\usepackage{cancel}

\providecommand{\abs}[1]{\left|#1\right|}

\providecommand{\ket}[1]{|#1\rangle}

\providecommand{\proj}[2]{|#1\rangle \! \langle#2|} 
\providecommand{\mean}[3]{\langle#1|#2|#3\rangle} 
\newcommand{\di}{\textrm{d}}

\newcommand{\deff}{{\, \vcentcolon = \,}}

\begin{document}

\title{Inequality-free proof of Bell's theorem}

\author{Andrea Aiello}
\email{andrea.aiello@mpl.mpg.de} 
\affiliation{Max Planck Institute for the Science of Light, Staudtstrasse 2, 91058
Erlangen, Germany}


\date{\today}

\begin{abstract}
Bell's theorem supposedly demonstrates an irreconcilable conflict between quantum mechanics and local, realistic hidden variable theories. 
Most proofs of Bell's theorem, are based on inequalities. In this paper we present an alternative proof which does not involve inequalities, but only a direct comparison between correlation functions calculated using quantum mechanics on the one hand, and those calculated according to local realistic hidden-variable theories on the other.
Our proof is based on a physically motivated use of Fourier series for periodic functions, and confirms that local realistic hidden-variable theories are incompatible with quantum mechanics.
\end{abstract}

\maketitle

\section{Introduction}
July next year will mark 100 years since Werner Heisenberg's memorable trip to the island of Helgoland, where he developed the first modern formulation of quantum  mechanics (QM), the so-called ``matrix mechanics'' \cite{Cassidy1993,WeinbergII}. Since then,  quantum mechanics has been developed and applied with great success in virtually all branches of physics as well as in other scientific disciplines, such as chemistry, medicine and information science, just to name a few.
Despite such many successes, the interpretation of the foundations of quantum mechanics is still hotly debated (see \cite{DespagnatBook} for a classic presentation of fundamental issues in quantum mechanics, and \cite{sep-bell-theorem} for a more up-to-date account of Bell's theorem), and the question often still arises as to whether it is possible to formulate alternative theories, closer in spirit to classical physics, which would reproduce quantum-mechanical results.

At first it was Einstein, Podolsky and Rosen \cite{PhysRev.47.777} who hypothesized that quantum mechanics was not a complete theory and should be supplemented by additional variables (hidden variables, in modern parlance).
Such interpretation of quantum mechanics was afterwards questioned  by Von Neumann  who presented an alleged proof that quantum mechanics is incompatible with some hidden variable theories  \cite{vonNeumann2018}. However, Von Neumann's conclusion was plagued by a conceptual mistake that was later pointed out by Bell \cite{RevModPhys.38.447}. Eventually, the latter managed to derive an inequality that would be satisfied by any local-realistic hidden variable theory, but which would be violated by quantum mechanics \cite{Bell1964}.
Since then, many variants of this celebrated inequality have been formulated, the most popular of which  is perhaps the CHSH (Clauser, Horne, Shimony et Holt) inequality  \cite{PhysRevLett.23.880}.
The rest is recent history and a fairly detailed and up-to-date reviews of Bell-inspired inequalities \cite{Ballentine},   can be found in \cite{RevModPhys.86.419,sep-bell-theorem}. Some critical views on this topic are expressed in section XIII.3 of \cite{Bohm1986}, and in \cite{DeBaere1984,ADENIER2001,Cetto2020,Lambare2021,Czachor_2023}. 
It is worth noting that since the creation of quantum mechanics, other alternative theories have been developed, with greater or lesser success, even without invoking the existence of hidden variables. One of the best known is  the Koopman-Von Neumann theory \cite{koopman1931hamiltonian,neumann1932operatorenmethode,neumann1932zusatze}.

Since Stapp's pioneering work \cite{PhysRevD.3.1303}, several authors have addressed the question of
proving  the incompatibility of QM with LHV theories by dispensing with inequalities \cite{10.1119/1.16243,ZUKOWSKI1993290,PhysRevA.49.2231,PhysRevA.61.022114,PhysRevLett.93.230403}.  Another interesting approach to Bell's inequality can be found in  \cite{doi:10.1119/1.4823600}.

In this paper we present a conceptually simple inequality-free proof of the incompatibility  of QM with LHV theories, based on the Fourier series expansion of periodic functions. The choice of this technique is entirely dictated by the physics of the problem, which is the analysis of the standard EPR-like experiment with entangled photon pairs, in the version elaborated by Bohm and Aharonov \cite{PhysRev.108.1070}. 

This paper is structured as follows. In Sec. \ref{EPR-like}, first we briefly review  the EPR-like experiment with photon pairs and two separated linear-polarization analyzers. Then, we calculate the correlations of the  outcomes of these analyzers, as given by both QM and LHV theories. In Sec. \ref{test}, we compare such correlations using the Fouriers series of periodic functions as main tool. We find, indeed, an irreconcilable conflict between quantum mechanics and local, realistic hidden-variable theories.  We discuss our results and draw our conclusions in Sec. \ref{conclusions}. An appendix present some detailed calculations.

\section{The EPR-like experiment}\label{EPR-like}

\begin{figure}[ht!]
  \centering
  \includegraphics[scale=3,clip=false,width=0.9\columnwidth,trim = 0 0 0 0]{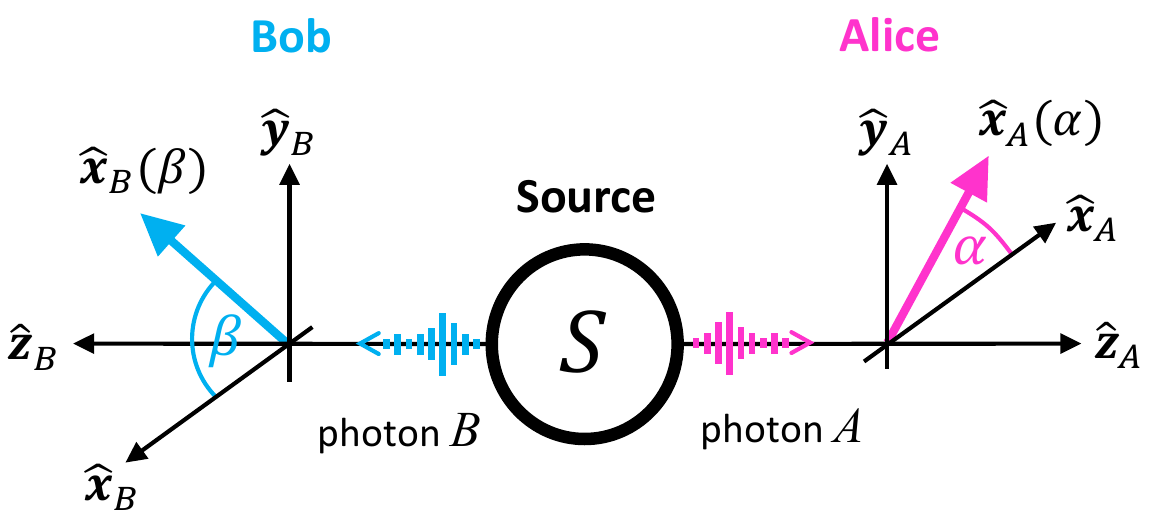}
  \caption{The source $S$ emits pairs of photons, say $A$ and $B$, in an entangled  polarization state. Photons $A$ and $B$ then propagates along the axes  $z_A$ and $z_B$, respectively, towards the observers Alice and Bob, which are located in two spatially separated regions. The unit vector $\hat{\bm{x}}_A(\alpha) = \hat{\bm{x}}_A \cos \alpha + \hat{\bm{y}}_A \sin \alpha$ defines the orientation of Alice's linear polarization analyzer, where the triad $\{ \hat{\bm{x}}_A, \hat{\bm{y}}_A, \hat{\bm{z}}_A\}$ forms a right-handed Cartesian reference frame for Alice's measurement apparatus. The same definitions hold, mutatis mutandis, for photon $B$ and Bob's  apparatus.}\label{fig1}
\end{figure}

Let us quickly illustrate the standard EPR-like experiment in the version elaborated by Bohm and Aharonov \cite{PhysRev.108.1070}, as sketched in Fig. \ref{fig1}. More complete descriptions can be found in \cite{Peres1995,Ballentine,CoTannoBookIII}.

The light source $S$ emits pair of photons, called  $A$ and  $B$, characterized by two independent degrees of freedom: the polarization and the path of propagation, all the other degrees of freedom being the same. 
After the emission,  photon $A$ propagates towards  Alice, the first observer, and photon $B$ goes to  Bob, the second observer. Both Alice and Bob  can test the polarization of the photons they receive, by means of a rotatable linear polarization analyzer. By hypothesis the two photons are emitted in the singlet polarization state
\begin{align}\label{a60}
\ket{\Psi} & =  \frac{1}{\sqrt{2}} \bigl(  \ket{x}_A \otimes \ket{y}_B -   \ket{y}_A \otimes \ket{x}_B \bigr) \nonumber \\[6pt]
& = \frac{1}{\sqrt{2}} \bigl(  \ket{x,y} -   \ket{y,x} \bigr),
\end{align}
where $\ket{x}$ and $ \ket{y}$ denote the state vector of a single photon (either $A$ or $B$), linearly polarized along the orthogonal axes $x$ and $y$, respectively. 
Here and hereafter when we write, for example,  $\ket{x}$,  we mean the state vector of a generic single photon linearly polarized along some axis $x$. On the other hand, when we write  $\ket{x}_A$, we intend the state vector of photon $A$ in an EPR-like experiment. The same convention applies to  observables and  operators. 

The action of a linear polarization analyzer oriented along the direction specified by the unit vector $\hat{\bm{x}}(\theta) = \hat{\bm{x}} \cos \theta + \hat{\bm{y}} \sin \theta$, with $\theta \in [0,\pi)$, can be described by means of the observable $Z(\theta)$, which in quantum mechanics is represented by the  Hermitian operator
\begin{align}\label{a70}
\hat{Z}(\theta) \deff \proj{x,\theta}{x, \theta} - \proj{y,\theta}{y, \theta},
\end{align}
where
\begin{equation}\label{a80}
\begin{split}
\ket{x,\theta} & = \cos \theta \ket{x} + \sin \theta \ket{y}, \\[6pt]
\ket{y,\theta} & = -\sin \theta \ket{x} + \cos \theta \ket{y},
\end{split}
\end{equation}
with $\ket{x} = \ket{x,0}, ~ \ket{y} = \ket{y,0}$. By construction, the eigenvalues of $\hat{Z}(\theta)$ are $+1$ and $-1$, associated with the eigenvectors $\ket{x,\theta}$ and $\ket{y,\theta}$, respectively. 

In this scenario, by the word ``experiment'' we denote the emission of a pair of photons by the source $S$, followed by a measurement of the observables ${Z}_A(\alpha)$ and ${Z}_B(\beta)$ by Alice and Bob, respectively.
The value of the product $Z_A(\alpha) Z_B(\beta)$ averaged over many repetitions of the experiment, gives the correlation of the outcomes of the two polarization analyzers.  Let us denote such \emph{measured} correlation with
\begin{align}\label{a90}
C^{M}(\alpha, \beta) = \overline{Z_A(\alpha) Z_B(\beta)}, 
\end{align}
where the superscript ``$M$'' stands for $M$easured, and the upper bar indicates average over many runs of the experiment. 

Physics places some constraints upon $C^M(\alpha, \beta)$. First, since $\alpha$ and $\beta$ are the angles along which Alice and Bob's linear polarization analyzers are oriented, then the angles $\alpha$ and $\alpha+\pi$, as well as the angles $\beta$ and $\beta+\pi$, must be  equivalent. This implies that 
\begin{align}\label{a100}
\mathrm{(C1)} & \qquad C^M(\alpha + n \pi, \beta + m \pi) = C^M(\alpha, \beta), \qquad
\end{align}
for all $ n,m \in \mathbb{Z}$. Second, since the two photons in each emitted pair are identical except for their polarization and propagation directions, then by swapping the orientation of the two polarization analyzers, the outcomes of the measurements cannot change.  Therefore, 
\begin{align}\label{a110}
\mathrm{(C2)} & \qquad \qquad \quad C^M(\beta , \alpha) = C^M(\alpha, \beta). \qquad \qquad  
\end{align}

\subsection{Correlation functions}

For the EPR-like experiment described above, quantum mechanics gives for the correlation of the outcomes of the two polarization analyzers, the expression
\begin{align}\label{a120}
C^Q(\alpha, \beta) & = \mean{\Psi}{\hat{Z}_A(\alpha) \otimes \hat{Z}_B(\beta)}{\Psi} \nonumber \\[6pt]
&  = -\cos \bigl[ 2( \alpha - \beta )\bigr],
\end{align}
where the superscript ``$Q$'' stands for $Q$uantum,  $\ket{\Psi}$ is given by \eqref{a60}, and the operators $\hat{Z}_A(\alpha), \hat{Z}_B(\beta)$ are defined by \eqref{a70}. It is straightforward to verify that $C^Q(\alpha, \beta)$ satisfies  both C1 and C2.

According to Bell \cite{Bell1964}, a local realistic hidden-variable theory characterized by a set of random parameters $\lambda$, predicts for the correlation of the outcomes of the two polarization analyzers, the function
\begin{align}\label{a130}
C^{H\!V}(\alpha, \beta) = \int_\Lambda \rho(\lambda)A(\alpha,\lambda) B(\beta,\lambda) \, \di \lambda,
\end{align}
where $\Lambda$ denotes the domain of the hidden variables $\lambda$, and $\rho(\lambda) \geq 0$, normalized to
\begin{align}\label{a140}
\int_\Lambda \rho(\lambda) \, \di \lambda = 1,
\end{align}
is the probability distribution of the parameters $\lambda \in \Lambda$, with the superscript ``$H\!V$'' standing for $H$idden $V$ariables. The two functions 
\begin{align}\label{a150}
A(\alpha,\lambda) = \pm1, \qquad B(\beta,\lambda)  = \pm 1,
\end{align}
supposedly determine the results of the outcomes of the measurements performed by Alice and Bob, respectively. 
Constraint C1 requires that both functions $A(\alpha,\lambda) $ and $B(\beta,\lambda) $ are periodic  with period $\pi$ with respect to the variables $\alpha$ and $\beta$, respectively: 
\begin{align}\label{a160}
A(\alpha + \pi,\lambda) = A(\alpha ,\lambda), \quad B(\beta + \pi,\lambda)  = B(\beta,\lambda).
\end{align}
More importantly, constraint C2 demands the two functions to be equal, that is 
\begin{align}\label{a170}
A(\theta,\lambda) = B(\theta ,\lambda) \deff F(\theta,\lambda), \qquad \forall \, \theta \in [0,\pi).
\end{align}
Therefore, we can rewrite \eqref{a130} as,
\begin{align}\label{a180}
C^{H \! V}(\alpha, \beta) = \int_\Lambda \rho(\lambda)F(\alpha,\lambda)F(\beta,\lambda) \, \di \lambda.
\end{align}

\section{The proof}\label{test}

At this point, the question we would like to answer is the following: is it possible to reproduce the quantum-mechanical correlation $C^Q(\alpha, \beta)$ by means of a local realistic hidden-variable theory? In other words, is it possible to find a function $F(\theta,\lambda)$ such that $C^{Q}(\alpha, \beta)  = C^{H \! V}(\alpha, \beta) $, that is
\begin{align}\label{a190}
-\cos \bigl[ 2( \alpha - \beta )\bigr] = \int_\Lambda \rho(\lambda)F(\alpha,\lambda)F(\beta,\lambda) \, \di \lambda \, ?
\end{align}

To answer this question, we first note that the quantities on either side of this equation are periodic functions with respect to  $\alpha$ and $\beta$, with period $\pi$. Therefore, we can write both sides of \eqref{a190} as Fourier series.
There are several different equivalent forms of the Fourier series, in this note we use the following one \cite{Appel}. Let $f(t): \mathbb{R} \to \mathbb{R}$ be a piecewise continuous real-valued function, periodic with period $T$. Then, the Fourier partial sum $f_N(t)$ is defined by
\begin{align}\label{i10}
f_N(t) = \sum_{n = -N}^N f_n \frac{e^{2 \pi i n t/T}}{\sqrt{T}},
\end{align}
where $N \in \mathbb{N}$, and the $n$-th Fourier coefficient $f_n$, is given by
\begin{align}\label{i20}
f_n = \int_0^T f(t) \frac{e^{-2 \pi i n t/T}}{\sqrt{T}} \, \di t,
\end{align}
with $n \in \mathbb{Z}$.

A direct calculation gives
\begin{align}\label{a200}
-\cos \bigl[ 2( \alpha - \beta )\bigr] = \sum_{n,m = -\infty}^\infty c_{nm} \, \frac{e^{2 i (n \alpha + m \beta)}}{\pi},
\end{align}
where
\begin{align}\label{a210}
c_{nm} = -\frac{\pi}{2} \bigl[ \delta(n,1)\delta(m,-1) + \delta(n,-1)\delta(m,1)  \bigr],
\end{align}
and $ \delta(i,j)$ denotes the Kronecker delta function defined by
\begin{align}\label{a220}
\delta(i,j) = \begin{cases}
0 &\text{if } i \neq j,   \\
1 &\text{if } i=j.   \end{cases}
\end{align}

The Fourier series of $F(\theta, \lambda)$, evaluated for a \emph{fixed} value of the parameters $\lambda$, is given by
\begin{align}\label{a230}
F(\theta, \lambda) =  \sum_{n = -\infty}^\infty f_{n}(\lambda) \, \frac{e^{2 i n \theta}}{\sqrt{\pi}},
\end{align}
where $ f_{n}(\lambda)$ are unknown $\lambda$-dependent coefficients calculated as
\begin{align}\label{a240}
f_{n}(\lambda) =  \int_0^\pi F(\theta,\lambda)  \frac{e^{-2 i n \theta}}{\sqrt{\pi}} \, \di \theta.
\end{align}
Substituting \eqref{a200} and \eqref{a230} in the left and right sides of \eqref{a190}, respectively, we obtain 
\begin{align}\label{a250}
\int_\Lambda \rho(\lambda)f_n(\lambda)f_m(\lambda) \, \di \lambda   = & \; -\frac{\pi}{2}  \delta(n,1)\delta(m,-1) \nonumber \\[6pt]
& \;  -\frac{\pi}{2}  \delta(n,-1)\delta(m,1).
\end{align}

We will now show that this equation cannot become an equality for any value of the Fourier series coefficients $f_n(\lambda)$ and $f_m(\lambda)$.
To this end, let us evaluate \eqref{a250} for $n=-m$ with $n \neq \pm 1$. In this case we obtain
\begin{align}\label{a260}
\int_\Lambda \rho(\lambda)f_n(\lambda)f_{-n}(\lambda) \, \di \lambda   =  0.
\end{align}
However, from \eqref{a240} it follows that since $F(\theta,\lambda) \in \mathbb{R}$, then 
\begin{align}\label{a270}
f_{-n}(\lambda) = f_{n}^*(\lambda) ,
\end{align}
where $z^*$ denotes the complex conjugate of the complex number $z$. Using this result we can rewrite \eqref{a260} as
\begin{align}\label{a280}
\int_\Lambda \rho(\lambda)\abs{f_n(\lambda)}^2 \, \di \lambda   =  0  .
\end{align}
Since $\rho(\lambda) \geq 0$ by definition, from \eqref{a280} it follows that
\begin{align}\label{a290}
f_n(\lambda)   =  0  , \qquad \forall \, n \neq \pm 1.
\end{align}
Substituting \eqref{a290} into \eqref{a230}, we obtain
\begin{align}\label{a230bis}
F(\theta, \lambda) =   f_{1}(\lambda) \, \frac{e^{2 i  \theta}}{\sqrt{\pi}} + f_{-1}(\lambda) \, \frac{e^{-2 i  \theta}}{\sqrt{\pi}}.
\end{align}
To determine $f_{\pm 1}(\lambda)$, we note that from $F(\theta, \lambda) = \pm 1$, it follows that
\begin{align}\label{a300}
\pi = & \; \sum_{n=-\infty}^{\infty} \abs{f_n(\lambda)}^2 \nonumber \\[6pt]
=  & \; 2 \abs{f_{\pm1}(\lambda)}^2,
\end{align}
where \eqref{a270} and \eqref{a290} have been used. Therefore, we can write
\begin{align}\label{a300bis}
f_{\pm 1}(\lambda) = \sqrt{\frac{\pi}{2}} \, e^{\pm i \varphi(\lambda)},
\end{align}
where
\begin{align}\label{phase}
 e^{\pm i \varphi(\lambda)} = \frac{f_{\pm 1}(\lambda)}{\abs{f_{\pm 1}(\lambda)}} \, .
\end{align}
Substituting \eqref{a300bis} into \eqref{a230bis}, we obtain 
\begin{align}\label{a230ter}
F(\theta, \lambda) = \sqrt{2}  \cos \bigl[\varphi(\lambda) + 2 \theta \bigr] .
\end{align}
This expression can be equal to $\pm 1$ at most for a finite number of values of $\theta$, determined by the equation $F(\theta, \lambda) =  \pm 1$, that is when $\varphi(\lambda) + 2 \theta = (2k+1)\pi/4$, with $k=0,1,2,3$.

Thus, we have demonstrated that it is impossible to reproduce the quantum-mechanical correlation \eqref{a120}, by means of a local realistic hidden-variable theory. This is the main result of our work.

\section{Discussion and conclusions}\label{conclusions}

The purpose of this paper was to demonstrate, without deriving any kind of inequality, that local realistic hidden-variable theories are incompatible with quantum mechanics. Our main result is that it is sufficient to verify, according to \eqref{a90} and \eqref{a120}, that $C^M(\alpha,\beta) = C^Q(\alpha,\beta) = -\cos \bigl[ 2( \alpha - \beta )\bigr]$, to rule out  hidden variable theories in the description of an EPR-like experiment. This implies that it is not necessary to take linear combinations of correlation functions that cannot be measured separately and simultaneously, as is believed to be the case with Bell's inequalities, as pointed out by Peres \cite{Peres1995}.

Within the present approach to Bell's theorem, an open question is that of quantum nonlocality \cite{Referee}.
As a matter of fact, violation of Bell's inequality (in any of its forms), can be used as a quantitative measure of quantum nonlocality  \cite{RevModPhys.86.419}. In recent years, this measure has found numerous applications in quantum information such as, for example, device-independent quantum key distribution \cite{RamonaWolf}.
This topic, per se, is very interesting and wide-ranging, so it cannot be discussed briefly here, but deserves a separate paper. Here we only briefly note that in the case of two-qubit bipartite entanglement that we studied in the present work, the number of Fourier coefficients of the correlation functions, could be considered as an indicator of nonlocality. It is so because we know that the quantum cosine-like correlation function $ C^Q(\alpha,\beta) = -\cos \bigl[ 2( \alpha - \beta )\bigr]$ can yield a maximum violation of Bell's inequality. The (qualitative) reasoning goes as follows:   the right side of \eqref{a200} is, in fact, a Schmidt decomposition \cite{10.1119/1.17904} of the function $-\cos \bigl[ 2( \alpha - \beta )\bigr] $. This can be easily seen by rewriting
\begin{align}\label{a200bis}
-\cos \bigl[ 2( \alpha - \beta )\bigr] = \sum_{n= \pm 1}\sqrt{\lambda_n} \, u_n(\alpha) v_n (\beta),
\end{align}
where  $\lambda_n \deff \pi/4$, and
\begin{equation}\label{a200ter}
\begin{split}
u_n(\alpha) &  \deff \exp(2 i n \alpha)/\sqrt{\pi},   \\[6pt]
v_n(\beta) & \deff -\exp(-2 i n \beta)/\sqrt{\pi}.
\end{split}
\end{equation}
 From $\lambda_1 = \lambda_{-1}$, it follows that the quantum correlation function $-\cos \bigl[ 2( \alpha - \beta )\bigr]$ is, from a mathematical point of view, maximally entangled with respect to the variables $\alpha$ and $\beta$. The Fourier series of any other correlation function which is not a simple cosine function,  will necessarily contain more than two terms. We can illustrate this point by taking as an example the simple hidden-variable model  proposed by Aspect in  \cite{Aspect1982}. In such a model, 
\begin{align}\label{p80}
A(\theta,\lambda) = \left\{
                                          \begin{array}{ll}
                                            +1, & \; \text{if} \; \; \cos[2(\theta - \lambda)] \geq 0, \\[6pt]
                                            -1, & \; \text{if} \; \; \cos[2(\theta - \lambda)] < 0,
                                          \end{array}
                                        \right.
\end{align}
and $B(\theta,\lambda) = - A(\theta,\lambda)$, with $\theta \in [0,\pi)$, and $\lambda \in \mathbb{R}$ uniformly distributed between $\lambda = 0$ and  $\lambda = \pi$, that is
\begin{align}\label{p90}
\rho(\lambda) = \frac{1}{\pi}, \qquad \lambda \in [0,\pi].
\end{align}
This choice yields the correlation function 
\begin{align}\label{p100}
C^{HV}(\alpha,\beta) & = \frac{1}{\pi}  \int_0^\pi A(\alpha,\lambda) B(\beta,\lambda) \, \di \lambda \nonumber \\[6pt]
& = \left\{
\begin{array}{cl}
 -1+ \frac{4 (\alpha - \beta) }{\pi }, & 0\leq \alpha - \beta <\frac{\pi }{2}, \\[6pt]
 3-\frac{4 (\alpha - \beta) }{\pi }, & \frac{\pi }{2}\leq \alpha - \beta <\pi. 
\end{array} \right.
\end{align}
It is not difficult to show that in this case 
\begin{align}\label{p110}
C^{HV}(\alpha,\beta) = \sum_{n= -\infty}^\infty \sqrt{\lambda_n} \, u_n(\alpha) v_n (\beta),
\end{align}
where $u_n(\alpha)$ and  $v_n (\beta)$ are still given by \eqref{a200ter}, and 
\begin{align}\label{p120}
 \sqrt{\lambda_n} = \left\{
\begin{array}{cl}
 0, & n \quad \text{even}, \\[6pt]
 \frac{4}{n^2 \pi^2 }, & n \quad \text{odd}. 
\end{array} \right.
\end{align}
Therefore, the Fourier series \eqref{p110} contains an infinite number of terms, as opposed to the only two terms of \eqref{a200bis}.

In conclusions, we  remark that we do not wish to contribute here the many pseudo-philosophical debates that often arise in discussions about the foundations of quantum mechanics. 
We simply emphasize that our position on the various, exotic interpretations of quantum mechanics and the alleged paradoxes, is of great skepticism, and is very close to that expressed, for example, by Sidney Coleman \cite{coleman2020sidney}, and Berthold-Georg Englert \cite{Englert2013}.
In particular, following Coleman, we believe that it is meaningless to try to interpret a new theory (quantum mechanics) in terms of an old theory (classical mechanics) \cite{coleman2020sidney}:
\begin{quote}
``\textit{The thing you want to do is
not to interpret the new theory in terms of the old, but
the old theory in terms of the new.}''
\end{quote}

\appendix

\section{About the function $F(\theta,\lambda)$}

It is interesting to note that the coefficients $f_n(\lambda)$ can be calculated, at least formally, up to a point because we know that whatever the functional dependence on $\lambda$ is, the function $F(\theta,\lambda)$ is always equal to $\pm 1$.
 This implies that for any given \emph{fixed} set of values of the parameters $\lambda$, $F(\theta,\lambda)$ is a periodic piecewise constant function with respect to the variable $\theta$, of period $\pi$ (in mathematical parlance, $F(\theta,\lambda)$ is a so-called \emph{simple function} \cite{RudinPrinciples}). Therefore,   $F(\theta,\lambda)$  is completely determined by a partition
\begin{align}\label{a310}
P(\lambda): \quad 0 = \theta_0 < \theta_1 < \ldots < \theta_{K} = \pi,
\end{align}
of the interval $[0,\pi)$, where $\theta_k = \theta_k(\lambda)$ and $K = K(\lambda) \in \mathbb{N}$, and by the sequence
\begin{align}\label{a320}
S(\lambda) = (s_1, s_2, \ldots, s_K), 
\end{align}
where $s_k = s_k(\lambda)$ is such that
\begin{align}\label{a330}
s_k = \pm 1,  \qquad (k=1, \ldots, K),
\end{align}
and 
\begin{align}\label{a335}
 s_{p} + s_{p+1} = 0, \quad (p=1, \ldots, K-1).
\end{align}
An  example of a possible $F(\theta,\lambda)$ in the interval $0 \leq \theta < \pi$, is shown in Fig. \ref{fig2} below.
\begin{figure}[h!]
  \centering
  \includegraphics[scale=3,clip=false,width=0.9\columnwidth,trim = 0 0 0 0]{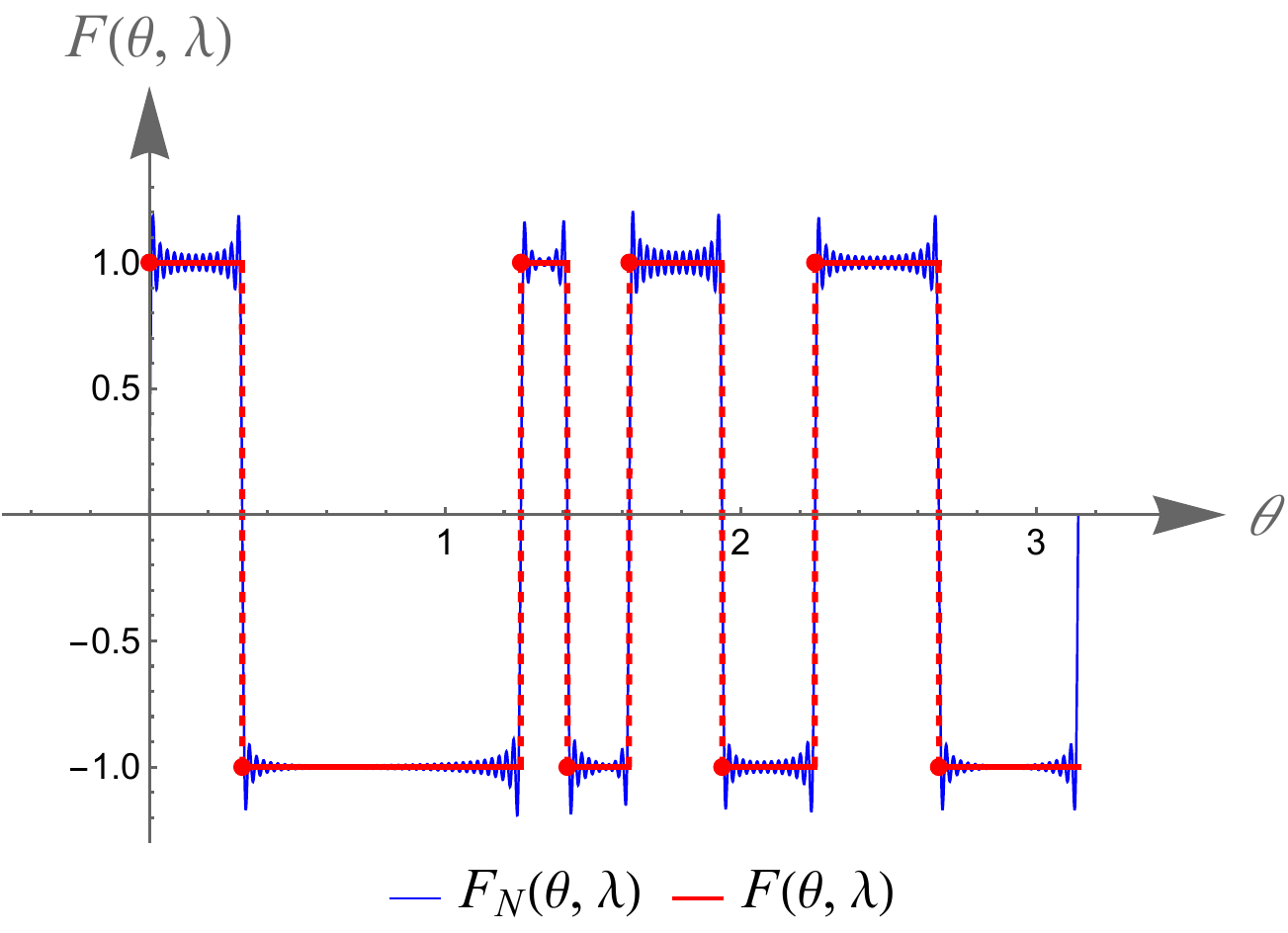}
  \caption{Plot of a possible realization of $F(\theta,\lambda)$ for $\lambda$ fixed, with $K=8$,  in the interval $\theta \in [0,\pi)$. In this figure $(s_1,  \ldots, s_8) = (1,-1,1,-1,1,-1,1,-1)$. The orange discontinuous line is the plot of $F(\theta,\lambda)$. The continuous blue line is the plot of $F_N(\theta,\lambda)$, which is the Fourier partial sum defined by \eqref{i10}, with $N = 128$ .}\label{fig2}
\end{figure}
\\
The partition $P(\lambda)$ determines a sequence of sub-intervals $I(\lambda) = (I_1, I_2, \ldots, I_K)$, such that $I_k = I_k(\theta, \lambda)$, and 
\begin{align}\label{a340}
I_k \deff [\theta_{k-1}, \theta_k) = \{ \theta \in \mathbb{R} ~|~ \theta_{k-1} \leq \theta < \theta_k \}, 
\end{align}
with $k=1, \ldots, K$, and $\theta_0 =0, \, \theta_K = \pi$, by definition. The length of $I_k$ is $l_k = \theta_k - \theta_{k-1}$. 
Note that the number $K$ of the sub-intervals $I_k$ in $P(\lambda)$ is determined by $\lambda$, so that $K = K(\lambda)$, because  different values of $\lambda$ yield different partitions of the same interval $[0,\pi)$.

At this point, if with $\bm{1}_{k}(\theta,\lambda) $ we denote the indicator function of the interval $I_k$, that is
\begin{align}\label{a350}
\bm{1}_{k}(\theta,\lambda) = \left\{
                       \begin{array}{ll}
                         1, & \hbox{for $\theta \in I_k$,} \\[6pt]
                         0, & \hbox{for $\theta \not\in I_k$,}
                       \end{array}
                     \right.
\end{align}
then, we can rewrite $F(\theta,\lambda)$ as
\begin{align}\label{a360}
F(\theta,\lambda) & = \sum_{k = 1}^{K} s_k(\lambda) \, \bm{1}_{k}(\theta,\lambda) \nonumber \\[6pt]
& = s_1(\lambda)\sum_{k = 1}^{K} (-1)^{k+1} \, \bm{1}_{k}(\theta,\lambda),
\end{align}
where $s_1(\lambda) = \pm 1$, and \eqref{a335} has been used.
Using this formula we can formally calculate the Fourier coefficients \eqref{a240}, obtaining
\begin{align}\label{a370}
 f_{n}(\lambda) = & \;  s_1(\lambda) \sum_{k = 1}^{K} (-1)^{k+1}  \int_0^\pi \, \bm{1}_{k}(\theta,\lambda) \frac{e^{-2 i n \theta}}{\sqrt{\pi}} \, \di \theta \nonumber \\[6pt]
= & \; s_1(\lambda) \sum_{k = 1}^{K} (-1)^{k+1} \int_{\theta_{k-1}}^{\theta_{k}} \, \frac{e^{-2 i n \theta}}{\sqrt{\pi}} \, \di \theta \nonumber \\[6pt]
= & \; i \, s_1(\lambda)  \sum_{k = 1}^{K} (-1)^{k+1} \frac{e^{- 2 i n \theta_{k}} - e^{- 2  n \theta_{k-1}}}{2  n \sqrt{\pi} }.
\end{align}
For $n=0$ this equation gives
\begin{align}\label{a380}
 f_{0}(\lambda) =  - \frac{s_1(\lambda)}{\sqrt{\pi}} \left[\pi(-1)^{K} + 2 \sum_{k = 1}^{K-1} (-1)^{k} \theta_k \right] ,
\end{align}
while for $n \neq 0$ we find
\begin{align}\label{a390}
 f_{n}(\lambda) = & \;  \frac{s_1(\lambda)}{2  i  n \sqrt{\pi}} \biggl[ 1 + \pi(-1)^{K} \nonumber \\[6pt]
&  + 2 \sum_{k = 1}^{K-1} (-1)^{k} \exp(- 2 i n \theta_k)  \biggr] ,
\end{align}
where $s_1(\lambda) = \pm 1$.

\end{document}